\documentstyle[pre,aps]{revtex}
\begin{document}

\draft

\def\theequation{\arabic{equation}}
\twocolumn[\hsize\textwidth\columnwidth\hsize\csname
@twocolumnfalse\endcsname

\title{A geometrical particle model for anyons.}
\author{
Armen Nersessian$^{1,2}$ and Eduardo Ramos$^3$}
\address{$^1$Bogoliubov Laboratory of Theoretical Physics,
Joint Institute for Nuclear Research,
141980 Dubna, Russia.\\
$^2$Department of Theoretical Physics, Yerevan State University,
A. Manoukian St., 5, Yerevan, 375012 Armenia  \\
$^3$ Dept. de F\'{\i}sica Te\'orica, C-XI
Universidad Aut\'onoma de Madrid
 Ciudad Universitaria de Cantoblanco 28049 Madrid, Spain}

\date{\today}
\maketitle
\begin{abstract}
We consider the simplest geometrical particle model associated with
light-like curves in $(2+1)$-dimensions. The action is proportional to
the pseudo-arc length of the particle's path.  We show that under
quantization it yields the $(2+1)$-dimensional anyonic field equation
supplemented with a Majorana-like relation on mass and spin, i.e.,
mass $\times$ spin $=\alpha^2$, with $\alpha$ the coupling constant in 
front of the action.
\end{abstract}
]

The search for lagrangians describing spinning particles, both massive
and massless, has a long history.  The conventional approach is based
on an extension of Minkowski space-time by auxiliary grassmann
variables which, after quantization, provide the extra degrees of
freedom required for integer or half-integer spin.  In contrast, in
planar physics spin is not quantized and particle states can be
anyonic, i.e., their spin can take any real value.  Thus, in the
general case, it is necessary to provide the initial classical model
with extra bosonic variables.  An interesting, and by now
well-explored, possibility is to supply those extra degrees of freedom
by lagrangians depending on higher order geometrical invariants.  The
requirements of Poincar\'e and reparametrization invariance restrict
the admissible set of such lagrangians to the ones depending only on
curvature and torsion.  An intensive study in this direction ,
performed in the late eighties by several authors (see \cite{misha}
and refs therein), was inspired by the remarkable work of Polyakov
\cite{polakov} on the $CP^1$ model minimally coupled to a Chern-Simons
gauge field.  An evaluation of the effective action for the charged
solitonic excitation yielded as a result
\begin{equation}
{\cal S}_{eff} =\int (m + {\pi\over 2\theta}\tau)|d{\bf x}|,
\quad |\dot{\bf x}|\neq 0
\label{actions}
\end{equation}
with $\tau$ the torsion of the particle's world-line, and $\theta$ the
gauge coupling strength.

However, on $(2+1)$-dimensional space one can consider actions of
other sort defined on light-like (or null) curves by
\begin{equation}
{\cal S}_0 =\int {\cal L}({\kappa})d{\sigma},\quad
d\sigma=|{\ddot{\bf x}}|^{1/2}dt,\quad {\dot{\bf x}}^2=0,
\end{equation}
with  $\kappa=|d^3{\bf x}/(d\sigma)^3|^2$ being the analog of the
torsion for null paths.

In a previous paper \cite{rn} we considered such an action, with a
constant ${\cal L}=2\alpha $, on $(3+1)$ dimensions. It was found that
upon quantization this system describes massive particles of arbitrary
integer or half-integer spin, and that its mass $m$ and spin $s$
satisfy the Majorana relation,  $ms=\alpha^2$. In this paper we study
the same system but immersed in $(2+1)-$dimensional space-time.
Explicitly our action for null paths is given by
\begin{equation}
{\cal S}_0 =2\alpha\int d\sigma .
\label{action0}
\end{equation}

We will show in the following that the $(2+1)$-dimensional case shares
all the nice properties of the $(3+1)$-dimensional system, with the
only exception that its spin can take arbitrary real values.  That is,
the system is essentially anyonic.

The essential tool to understand the geometry of null curves in
$(2+1)$ is supplied by their Frenet equations.  A moving frame adapted
to the description of light-like curves is given by
\begin{equation}
({\bf e}_{\pm}, {\bf e}_1):\;
{\bf e}_{\pm}{\bf e}_{\pm}={\bf e}_\pm{\bf e}_1 =0,
\;{\bf e}_+{\bf e}_-=1,
\; {\bf e}^2_1 =-1,
\label{mf}\end{equation}
with vector product
\begin{equation}
{\bf e}_+\times {\bf e}_-={\bf e}_1,\quad
{\bf e}_\pm\times {\bf e}_1=\pm{\bf e}_\pm.
\end{equation}

With these conventions the pseudo-arc length $\sigma$, and the
curvature $\kappa$, are defined via the Frenet equations by:
\begin{equation}
{d{\bf x}\over d\sigma}={\bf e}_+,
\; {d{\bf e}_+\over d\sigma}={\bf e}_1,
\;{d{\bf e}_1\over d\sigma}=\kappa {\bf e}_+ +{\bf e}_-,
\; {d{\bf e}_-\over d\sigma}=\kappa{\bf e}_1.
\label{ff}\end{equation}
Therefore,
\begin{equation}
{\dot\sigma}=\sqrt{-{\dot{\bf e}}^2_+},
\quad 2\kappa=\left({d{\bf e}_1\over d\sigma}\right)^2
\label{pa}\end{equation}

One may now rewrite our initial lagrangian in an equivalent first
order form:
\begin{equation}
 L=2\alpha{\sqrt{-\bf{\dot e_+}^2}}+
{\bf p}({\bf{\dot x }}-{\sqrt{-\bf{\dot e_+}^2}}{\bf e}_+)-
\lambda {\bf e}_+^2
\end{equation}
where ${\bf x}, {\bf p},{\bf e}_+,\lambda$ are independent variables.

By direct application of Dirac's formalism for singular lagrangians,
one obtains the resulting Hamiltonian system (in the pseudo-arc
gauge, ${\dot\sigma}=1$):
\begin{equation}
\begin{array}{c}
\Omega=d{\bf p}\wedge d{\bf x}+
 d{\bf p}_+\wedge d{\bf e}_+ ,\\
{\cal H}=(2\alpha)^{-1}\left({\bf p}_+^2 +({\bf p}{\bf e}_+-2\alpha)^2
+{\bf p}^2{\bf e}_+^2\right),
\end{array}
\end{equation}
supplied with  the constraints
\begin{equation}
\left\{
\begin{array}{c}
\phi_0\equiv {\bf p}_+^2 +({\bf p}{\bf e}_+-2\alpha)^2\approx 0,\\
\phi_1\equiv{\bf e}_+^2\approx 0,\\
\phi_2={\bf p}_+{\bf e}_+\approx 0,\\
\phi_3\equiv{\bf p}{\bf e}_+ -\alpha\approx 0,\\
\phi_4\equiv {\bf p}{\bf p}_+\approx 0.\\
\end{array}
\right.
\label{sc}
\end{equation}

Consistency of the equations of motion with those of Frenet requires
that
\begin{equation}
{\bf e}_1=-{\bf p}_+/\alpha,\quad {\bf e}_-={\bf p}/\alpha+\kappa{\bf e}_+,
\quad \kappa=-{\bf p}^2/2\alpha^2.
\end{equation}

The reduced phase space is {\it six\/}-dimensional because there is
only one first-class constraint, ${\cal H/\sigma}$.  In order to
obtain one-particle states one should eliminate the extra degrees of
freedom. This is naturally achieved by imposing the mass-shell
condition
\begin{equation}
{\bf p}^2= m^2,
\label{m}\end{equation}
which is a constant of motion for our system \cite{one}.

The generators of Lorentz transformations read:
\begin{equation}
 {\bf J}={\bf p}\times {\bf x} +
{\bf p}_+\times {\bf e}_+= {\bf p}\times {\bf x} + \alpha{\bf e}_+ .
\end{equation}
{}From this expression and the constraint $\phi_3$ one directly
obtains that
\begin{equation}
{\bf p}{\bf J}= \alpha^2,
\end{equation}
which is tantamount of saying that there is a Majorana-type relation
between mass and spin given by
\begin{equation}
ms=\alpha^2.
\end{equation}

If one introduces now the coordinates
\begin{equation}
{\bf X}\equiv {\bf x} -\frac{\alpha}{{\bf p}^2}{\bf p}_+,\quad
{\bf E}_+ \equiv {\bf e}_+ -\frac{\alpha}{{\bf p}^2}{\bf p},
\end{equation}
the symplectic structure $\Omega$ takes the simple form (on the
constrained surface)
\begin{equation}
\Omega=d{\bf p}\wedge d{\bf X} + d{\bf p}_+\wedge d{\bf E}_+\;\;,
\end{equation}
while the constraints read
\begin{equation}
\begin{array}{c}
{\bf p}{\bf p}_+= 0,\quad {\bf p}{\bf E}_+= 0,\\
{\bf p}_+^2 +\alpha^2= 0,\quad
{\bf p}_+{\bf E}_+= 0,\quad
{\bf E}^2_+ + {\alpha^2/{\bf p}^2}= 0.
\end{array}
\end{equation}
The equations of motion for ${\bf X}$, ${\bf E}_+$, ${\bf p}_+$, ${\bf
p}$ are
\begin{equation}
{d^2{\bf X}\over (d\sigma)^2}=0,\quad {d{\bf p}\over d\sigma}=0,
\end{equation}
and
\begin{equation}
{d^2{\bf E}_+\over (d\sigma)^2} =2\kappa {\bf E}_+,\quad
{d^2{\bf p}_+\over(d\sigma)^2} =2\kappa {\bf p}_+.
\end{equation}
Therefore ${\bf X}$ may be interpreted as the free coordinate
associated with the particle, while ${\bf E}_+$  is the responsible of
the {\it Zitterbewegung\/} associated with the spin degrees of
freedom.

It will show convenient to introduce the complex coordinate
\begin{equation}
{\bf z}={\bf p}_+ +i
\sqrt{{\bf p^2}}{\bf E}_+.
\end{equation}
In terms of this coordinate the constraints are given by
\begin{equation}
{\bf z}^2= 0, \quad {\bf z}{\bf{\bar z}} + 2\alpha^2= 0,\quad
{\rm and}\quad {\bf p}{\bf z}= 0,
\label{const}\end{equation}
while symplectic one-form $\theta$ and the Lorentz generators ${\bf
J}$ read
\begin{equation}
\theta={\bf p}d{\bf X}+
\frac{\imath({\bf z}d{\bf{\bar z}}- {\bf{\bar z}}d{\bf z})}{4m},
\quad
{\bf J}={\bf p}\times{\bf X}+
\frac{\imath}{2 m}{\bf z}\times{\bf {\bar z}}.
\label{ajz}\end{equation}

One may resolve the constraints by noticing that the first two of them
impose that ${\bf z}$ may be written in terms of a single complex
parameter $\omega$ as
\begin{equation}
{\bf z} = {\alpha\over \imath (\omega-\bar\omega)}
\left( 1+\omega^2 , 1 -\omega^2 , 2\omega \right).
\end{equation}
{}From the remaining constraint ${\bf pz}=0$ follows
that
\begin{equation}
\omega ={i p_2 \pm m\over p_0 + p_1}.
\end{equation}
So one can finally write the symplectic one-form $\theta$ and Lorentz
generators ${\bf J}$ solely in terms of ${\bf X}$ and ${\bf p}$.
It is a straightforward analytical computation to check that on the
${\bf p}^2 =m^2$ hypersurface the induced symplectic two-form
coincides with the one of the standard ``minimal covariant model'' for 
anyons \cite{jakiw}, which is given by the Dirac's monopole-like
two-form
\begin{equation}
d\theta= d{\bf p}\wedge d{\bf X}\pm
s\frac{\epsilon_{ijk}p^idp^j\wedge dp^k}{2m^3}.
\end{equation}
As it is well known, from the fact that this two-form is globally
defined it follows that the spin $s$ is not quantized.  For a detailed
explanation of this, as well as other aspect of the quantum theory of
anyons, we refer the interested reader to \cite{jakiw,lob}.\\
{\large Acknowledgements.}The work of A.N. has been partially
supported by grants INTAS 93-127-ext, INTAS 96-538,  and INTAS-RFBR
95-0829.

\end{document}